\documentclass{article}
\usepackage{mrs2005,epsfig}
\begin{document} 
\title{GDDS HIGHLIGHTS: GALAXY EVOLUTION REVEALED}

\author{DAVID CRAMPTON}
\affil{Dominion Astrophysical Observatory, HIA, NRC Canada, Victoria, BC, Canada}
\author{The GDDS Team: R. Abraham, R. Carlberg, H-W Chen, K. Glazebrook,
I. Hook, I. J{\o}rgensen, S. Juneau, D. LeBorgne, P. McCarthy, R. Marzke,
R. Murowinski, K. Roth, S. Savaglio }

\begin{abstract} 
The Gemini Deep Deep Survey, GDDS, produced several significant
results relating to the evolution of galaxies. All of these results
are consistent with the ``downsizing" concept of 
galaxy formation and evolution, i.e., that the active periods of star formation moved 
progressively from very massive galaxies at high redshift to much lower 
mass galaxies at the present epoch. Spectra of massive red galaxies at 
z $\sim$ 1.7 demonstrates that they contain old stellar populations and 
hence must have formed their stars in the first $\sim 3$Gyr of cosmic 
history; indicators of star formation activity show that the star 
formation rate in the most massive galaxies was much higher at $z = 2$ 
than today, that the activity in intermediate mass galaxies peaked near
$z \sim 1.5$, while, since $z \sim 1$ the activity is primarily confined to lower mass 
galaxies. The GDDS also uncovered a relatively high percentage of post-starburst galaxies
at $z \sim 1$, a result that is anticipated given all the activity seen at higher 
redshifts. Measurements of the strengths of metal lines of a 
subsample of the GDDS and CFRS galaxies at $z \sim 0.7$ reveal 
that, at a given mass, they had lower metallicities than at present. 
The evolution in the mass-metallicity relation is consistent with a 
model in which star formation lasts longest in less massive galaxies, 
again an expected result in the downsizing scenario. 
\end{abstract} 
 
\section{Introduction} 

The formation and evolution of galaxies has been one of the primary topics 
in cosmology ever since their diverse appearance was first observed. 
Naive interpretations of the Hubble sequence of galaxies and of the
hierarchical nature of dark matter halos in CDM models naturally lead 
to scenarios in which smaller galaxies, predominantly spirals, form first 
and then merge to form the massive spheroidal systems. The variation of the star 
formation rate density as a function of redshift (Lilly et al. 1996,  
Madau et al.1996) shows that almost half of the stellar mass in galaxies was formed 
between $1 < z < 2$ so that it is obviously vital to study this epoch to understand 
galaxy evolution. The GDDS survey (Abraham et al. 2004), through preselection criteria
applied to a deep infrared survey of galaxies, was specifically 
designed for this purpose. 

The photometric and spectroscopic results that emerged from 
the GDDS galaxy sample reveal a very consistent picture; one 
in which star formation activity was initially concentrated in very massive 
galaxies at very high redshift resulting in old, ``red and dead" galaxies 
at redshifts $z \sim 2$, and the activity has progressively 
moved to lower and lower mass systems. Here we summarize some of
the highlights that lead to this conclusion.

\section{Parameters of the GDDS Survey} 

The chief observational hurdle to overcome was: a) the extreme faintness
of the passively evolving galaxy population; b) the relatively featureless spectra
(since all strong features were redshifted out of the detector
passband); c) the very bright night sky emission. 
Before techniques like Nod and Shuffle (see below) were implemented, it
was extremely difficult to obtain sufficiently good spectra for redshift
determination in the range $1.4 < z < 2$ and so this became known
as the ``redshift desert". Given the observational challenges, transforming 
the redshift desert into the ``redshift
dessert" requires extremely long exposures.

The basic goals and parameters of the GDDS survey can be summarized as follows:

- Construct the largest mass-limited sample of galaxies in 
the $1 < z < 2$ range with no bias toward emission line systems. 
In fact, bias the sample toward early-type ``red and dead" galaxies 
using VRIzJHK colours. The sample was drawn from the Las Campanas 
IR imaging survey (McCarthy et al. 1999; Chen et al. 2002) and the 
GDDS subsample covered 121 square arcminutes in 4 different sightlines. 

- Go deep enough ($K_{Vega} < 20.6$) to pick up M* galaxies out to z=1.8. 
This requires up to 30 hour integrations with GMOS, the Gemini Multi-Object 
Spectrograph (Hook et al. 2003; Crampton \& Murowinski 2004) and implementation 
of a ``Nod \& Shuffle" mode (Cuillandre 1994, Glazebrook \& Bland-Hawthorn 2001, and
the appendix to Abraham et al. 2004)
to enable sky subtraction to 
0.1\% accuracy. The resulting relatively high signal-to-noise spectra that
were free of artifacts resulting from poor subtraction of the strong
night sky emission lines enabled redshift 
determination for a high perecentage of the original galaxy sample, even from
spectra of galaxies containing primarily 
old stars with no, or weak, emission features. Spectroscopic redshifts 
were derived for 221 galaxies, 163 of which are in the redshift interval $0.8 < z < 2 $. 

- Construct the mass function of galaxies and connect this to 
the star-formation history of the Universe (ideally, using only 
self-consistent data internal to the survey). Connect mass-assembly 
to the Hubble sequence.

The GDDS catalogue and the spectra are described in detail in Abraham et al. (2004) and
are all publicly available at http://www.ociw.edu/lcirs/gdds.html. Abraham et al. also 
describe the sampling strategy and how a weighting scheme was devised to deal
with the complications arising from the preselection criteria.

\section{Masses and Spectra of Galaxies in the Redshift Desert}

As outlined in Glazebrook et al. (2004), masses for galaxies in the GDDS 
sample were derived from their spectral energy distributions which in turn were
determined from $VIzK$ colours. As figures 1 and 3 of Glazebrook et al. 
show, high mass galaxies were abundant up to redshifts $z \sim 2$, 
and the decline in mass density falls much more slowly with 
redshift than predicted by standard semi-analytic models 
(e.g., Cole et al. 2000). Almost half of the mass density at 
$1.3 < z < 2$ is contained within galaxies with $I - K > 4$ 
and the spectra show that these are galaxies containing
populations of old stars rather than being reddened by interstellar 
absorption. Furthermore, HST ACS imaging of a representative 
subsample of these galaxies shows that more than 90\% have 
early-type morphology (Abraham et al. 2005). A representative spectrum
and direct image of one of these galaxies is shown in the middle panel of Figure 1.
The most massive 
galaxies are generally spectroscopically and morphologically 
early type systems. However, there are also lots of massive star-forming 
galaxies at $z > 1.4$ that presumably become all the post-starburst galaxies 
that are observed at $z \sim 1$ (see below).

\subsection{Formation redshifts and ages of galaxies with very evolved stellar populations}

A preliminary analysis by McCarthy et al. (2004) of the spectra of
galaxies at redshifts from 1.3 to 2.2 whose integrated light
is dominated by evolved stars shows that they must have formed
at very high redshifts. Conservative best-fit and minimum ages were derived
from both the spectra and the broad band 
colors. {\it {Minimum}} formation redshifts $>3.5$ are inferred
for several of the galaxies
and {\it{best-fit}} formation redshifts $z >5$  for 20\% of the objects.
These early-forming galaxies are major contributors 
to the stellar mass density at $1 < z < 2$ and are likely 
progenitors of a significant
fraction of present day massive elliptical galaxies. McCarthy et al.
argue that the mostly likely 
progenitors of these systems are higher redshift analogs of the massive 
starburst galaxies seen in the sub-mm with SCUBA and similar instruments,
and cannot be
descendants of the $z \sim 3$ Lyman break galaxies. 

Another interesting point is that the spectra are
best fit by models with solar or even higher metallicity indicating
that very significant metal enrichment occurred very early. The interstellar medium
of star-forming galaxies in this same redshift range is also 
apparently chemically evolved. A composite spectrum derived from a
subsample of 13 GDDS galaxies at $1.3 < z < 2$ 
by Savaglio et al. (2004) shows surprisingly strong interstellar lines,
much stronger than observed in local starbursts (Tremonti, 2003). An
example of the spectra of one of these active starburst galaxies at $z = 1.5$ is
shown in the top panel of Figure 1. Thus, both the interstellar medium and
the stars in these relatively high redshift galaxies have already been enriched
to approximately solar values. Rapid enrichment in these massive galaxies must
occur very early in the history of the Universe.

%
%
\begin{figure}  
\begin{center}
\epsfig{figure= 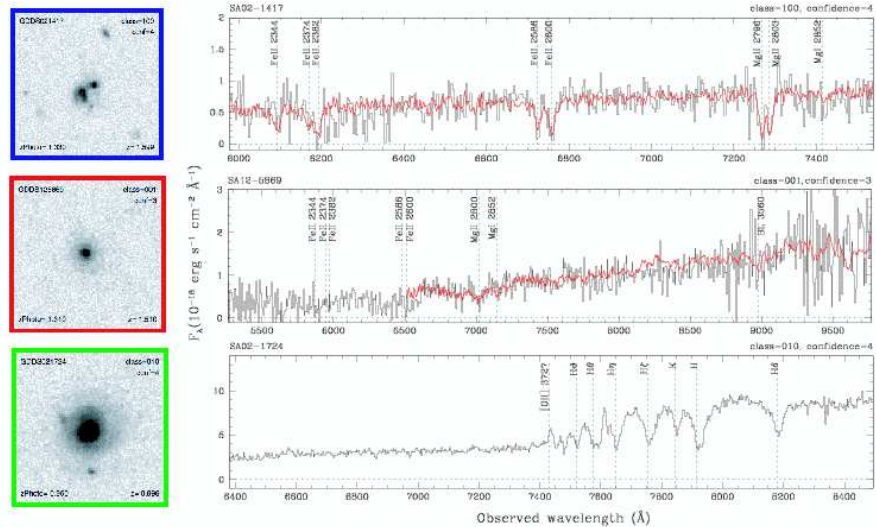,width=12.5cm}  
\end{center}
\caption{GMOS spectra and $HST ACS$ images representative of the GDDS sample.
Upper panel: A spectrum of a strongly star-forming galaxy with late-type morphology.
The flat spectrum in the rest-frame UV is a signature of active star formation while the
strength of the interstellar absorption features reveals that the metallicity must already be high
in this galaxy at $z = 1.6$. Middle panel: A spectrum of an old red and dead galaxy at
$z = 1.5$ with corresponding early-type morphology. Bottom panel: A spectrum and an image of a 
post-starburst galaxy at $z = 1$. Note the very strong Balmer series.
} 
\end{figure}
%
%
\begin{figure}  
\begin{center}
\epsfig{figure= 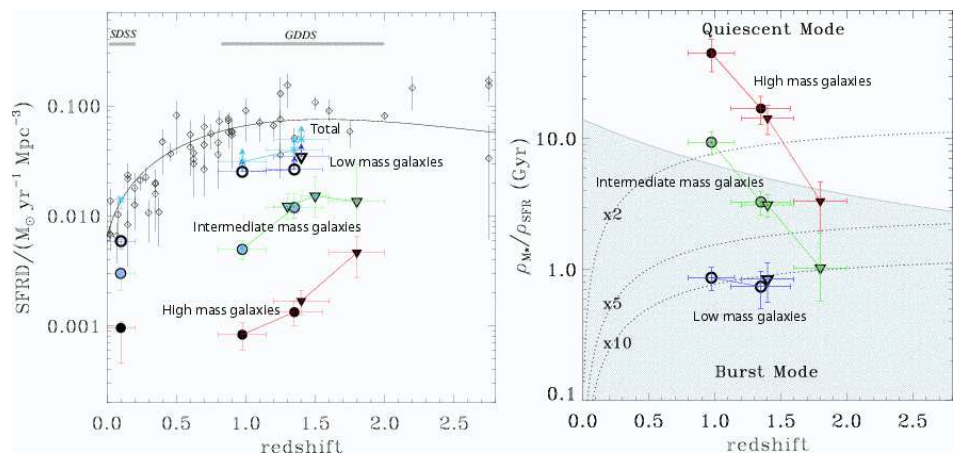,width=12.5cm}  
\end{center}
\caption{Left: The star formation rate density (SFRD) as a function of redshift. At the bottom of the panel,
the points show that the SFRD for GDDS high mass galaxies peaks at redshifts near $z \sim 2$ and declines
to present day values by $z \sim 1$. The SFRD for lower mass galaxies peaks at progressively lower
redshifts. Right: The characteristic timescale of the growth of stellar mass in galaxies as a function 
of redshift. By comparing the time required for the galaxies to assemble their mass at the observed star
formation rate with the age of the universe (gray curve) it is obvious that the stellar mass in high mass galaxies
built up rapidly before $z \sim 1.8$ and then these galaxies transitioned to a quiescent mode. Lower mass
galaxies do this at lower redshift. See Juneau et al. 2005 for details.
} 
\end{figure}
 
\subsection{Post-Starburst Activity at $z \sim 1$} 

Prominent among the GDDS spectra are several dramatic examples of post-starburst
galaxies, i.e., spectra with very strong Balmer absorption lines but weak or absent emission
features (see the bottom panel of Figure 1). By comparing the GDDS data with
that from the Sloan Digital Sky Survey, Le Borgne et al. (2005) were able to
quantify the dramatic decline in the abundance of post-starburst galaxies from
$z \sim 1.2$ to the present. Spectral synthesis models of the GDDS post-starburst
galaxies show that they must be the products of massive bursts of star formation
$1 - 2$ Gyr prior to the epoch of observation. The models show that these must have
been rapid mass-building events rather than a continual series of small events. The bright
star-forming galaxies in the GDDS sample at $z \sim 1.5$ are probably examples of this
activity.  The
most likely local counterparts of these post-starburst galaxies at $z \sim 1$
are massive ellipticals
in the field or in small groups.

\subsection{Star Formation Rates at $0.8 < z < 2$}

Another ubiquitous feature of the GDDS spectra is the presence of relatively
strong emission lines, much stronger than in lower redshift samples, e.g., CFRS.
[OII] emission lines, indicative of star formation activity, are, of course,
strong in the galaxies with late-type morphologies but in the GDDS sample
they are also frequently visible
even in spectra that appear to arise mostly from evolved populations. In addition,
higher ionization emission lines (e.g., [Ne III], [Ne V]) are evident in several
of the most active galaxies (for details, see Abraham et al. 2004). Juneau et al. (2005)
measured the strengths of the [OII] emission lines in all the GDDS spectra of
galaxies with $0.5 < z < 1.6$ and used these, along with measurements
of the rest-frame UV continuum (derived from photometry) for galaxies
with $1.2 < z < 2$ to derive star formation rates for the GDDS sample. As expected,
given the appearance of the spectra, the rates are much higher on average
than at $z = 0$. The availability of the mass estimates for the GDDS galaxies
allowed Juneau et al. to add another dimension, however. 
Figure 2, adapted from Juneau et al.,
demonstrates that the star formation rate history is a strong function of
the stellar mass in a galaxy. For the most massive galaxies, the star formation  rate was 
six times higher at $z \sim 2$ than at $z = 0$ and it declined rapidly between
$1 < z < 2$. The star formation activity in the lower mass galaxies peaked at
lower redshifts so that for systems with $9.0 < log(M_*/M_{sun}) < 10.2$ the star
formation activity peaks closer to $z \sim 1$ and then declines rapidly to
$z = 0$. Through a comparison of the star formation rate density to the
stellar mass density Juneau et al.  were able to demonstrate that star formation
in the most massive galaxies transitioned from a very active burst mode to
a quiescent mode near $z = 1.8$, whereas this transition occured later, at lower
redshifts, for less massive galaxies. Thus, measurements of the star formation
rate point to the same picture as was derived from analysis of the ages of the
populations from their spectra: the most massive galaxies formed early. In fact,
the most massive galaxies formed their stars in the first 3 Gyr of cosmic history.

All of the key results from the GDDS spectra and colours that are described
in this section strongly support the
``downsizing" concept for galaxy evolution: rapid star formation occurred very early
at high redshift
in the most massive systems and then subsided in these while the activity proceeded
to lower and lower mass systems with decreasing redshift.

\section{Redshift Evolution of the Galaxy Mass-Metallicity Relationship}

Savaglio et al. (2005) were able to measure metallicities for 56 galaxies
with $0.4 < z < 1.0$ for which masses could also be estimated from the GDDS and 
CFRS surveys. Careful attention was paid to minimizing systematic and possible aperture
effects as well as the effects of extinction. A nice mass-metallicity relationship
emerged for these galaxies at a mean redshift $z \sim 0.7$, a relationship that is
considerably tighter than the luminosity-metallicity relationship. Furthermore, the relation is 
displaced from that derived from the SDSS galaxies at $z \sim 0.1$ (Tremonti et al. 2004)
in the expected direction: 
present-day galaxies of a given mass have a higher metallicity. Evidence from $z \sim 2.3$
galaxies compiled by Shapley et al. (2004) indicates that this trend continues to higher
redshifts. Savaglio et al. show that all these observations can be fitted by a simple
redshift-dependent mass-metallicity relationship. Both masses and metallicities evolve 
more slowly in lower mass galaxies compared to their higher mass counterparts. The variation
can be explained by a simple closed-box model in which the e-folding time for star formation
is longer for less massive galaxies. Thus the picture that emerges from considerations
of the mass-metallicity relation also supports the ``downsizing" scenario.
An aside is that the mass-metallicity
relationship predicts that low metallicity galaxies such as the DLA systems all have masses
$M \sim10^{8.8}M_{sun}$. Hammer (elsewhere in these proceedings, and see Liang et al. 2005)
also reported detecting redshift evolution in the mass-metallicity 
relationship of galaxies at similar redshifts to those discussed by Savaglio et al.

\section{Summary} 
 
 The GDDS survey targeted $1 < z < 2$ galaxies, emphasizing those ``red and dead" systems containing
 old, evolved, stellar populations. As expected, the results demonstrate that UV-selected surveys are 
 not complete and miss important segments of the galaxy population. The GDDS spectra have sufficient
 signal-to-noise for these optically extremely faint galaxies that it is possible to distinguish the
 evolved populations from galaxies that are red because they are dusty. In addition, $HST ACS$ imaging
 confirms that almost all the spectroscopically ``red and dead" systems have morphologies characteristic
 of early-type galaxies.
 
 The abundance of massive galaxies declines at $z > 1$ but the decline is slower than that predicted by
 most models and there is still a significant population of very high mass galaxies at $z = 2$. The GDDS
 observations indicate that half the stellar mass in these galaxies was formed in very active bursts of star
 formation at $2 < z < 5+$. High redshift sub-mm ``SCUBA" sources (Chapman et al. 2005) 
 are the most plausible counterparts of 
 these active systems. At much lower redshifts, $z \sim 1$, there is a significant population of
 post-starburst galaxies that appear to be the descendants of the active star-forming galaxies at $1.4 < z < 2$.
 Measures of star formation activity in GDDS galaxies at various redshifts as a function of the stellar mass 
 complement these other results, demonstrating that the peak of the star formation activity began in the highest 
 mass galaxies at high redshifts and then progressively moved to lower mass galaxies at lower redshifts.
 This scenario, called ``downsizing" by Cowie et al. (1994), is thus supported by several results emerging
 from the GDDS survey. The redshift-evolution of a mass-metallicity relation that was detected in the GDDS and CFRS
 galaxies is also consistent with downsizing in that star formation appears to proceed more slowly, for a longer
 period of time, in less massive galaxies. Interestingly, a naive closed-box model in which the e-folding time for
 star formation is longer in less massive galaxies fits the observations quite well (Savaglio et al. 2005).
 
 In summary, observations of galaxies in the redshift desert give strong support for the downsizing scenario of
 galaxy evolution. Physical processes must make the formation of stars much more efficient in high mass systems 
 at earlier times than suggested by most theoretical models, and then turn it off quickly.

\vfill 
\end{document}